# Time as a Statistical Variable and Intrinsic Decoherence

Rodolfo Bonifacio

Dipartimento di Fisica, Università degli Studi di Milano

INFN and INFM, Sezione di Milano, Via Celoria, 16, 20133, Milano, Italy

We propose a novel approach to intrinsic decoherence without adding new assumptions to standard quantum mechanics. We generalize the Liouville equation just by requiring the dynamical semigroup property of time evolution and dropping the unitarity requirement. With no approximations and specific statistical assumptions we find a generalized Liouville equation which depends on two characteristic time $\tau_1$ and $\tau_2$ and reduces to the usual equation in the limit $\tau_1=\tau_2\to 0$. However, for $\tau_1$ and $\tau_2$ arbitrarily small but finite, our equation can be written as a finite difference equation which predicts state reduction to the diagonal form in the energy representation. The rate of decoherence becomes faster at the macroscopic limit as the energy scale of the system increases. In our approach the evolution time appears, a posteriori, as a *statistical variable* with a Poisson-$\Gamma$ function probability distribution *as if* time evolution would take place randomly at average intervals $\tau_2$ each evolution having a time width $\tau_1$. This view point is supported by the derivation of a generalized Tam Mandelstam inequality. The relation with previous work by Milburn, with laser and micromaser theory and many experimental testable examples are described. The agreement with recent experiments on damped Rabi oscillations is discussed.

**1. Introduction:**

The existence of coherent superposition of states is the basic reason for many paradoxical aspects of quantum mechanics. The evolution from a coherent superposition state to a statistical mixture is called decoherence. This is a central problem for measurement theory and for the classical limit of quantum mechanics at the macroscopic level. The Schroedinger cat, which can be in a superposition of states dead or alive, or a macroscopic particle which can be in a superposition of "here" and "there" are typical example of paradoxes whose interpretation is still controversial. A superposition state gives *non zero off-diagonal elements of the density operator*, which give rise to *quantum interference* and *non-classical correlation effects*. von Neuman postulated the reduction to the diagonal form as a result of a measurement. However, this reduction remains mysterious because it cannot be described by a unitary Hamiltonian evolution and does not clarify *how* and *when* state reduction takes place and what the underlying dynamical process is. Essentially two approaches to decoherence have been proposed: the most widely used [1,2] takes quantum mechanics as it is, appealing to dissipation due to the interaction of the system with the environment or, equivalently, with the measuring apparatus, tracing at the end, over the many degree of freedom of the environment. The decoherence of the system is usually described by Master Equations (ME) which are derived using reasonable statistical assumptions and *specific but arbitrary models to simulate the environment* and *its interaction with the system*. Furthermore, many approximations such as perturbative expansion, Markov approximation, etc. are used. From a conceptual view point, it appears rather peculiar that one has to invoke the environment, dissipation or the measuring apparatus to conclude that the cat must be dead or alive and a big particle must be here or there and not in a superposition of these possibilities with interference between them. It is like saying that the particle is going to be localized here as a result of is friction, and that the macroscopic limit is due to in the fact that the damping becomes stronger in this limit.

A minority of scientists look for a modification of *standard quantum mechanics* to include decoherence as something *intrinsic*, i.e., not related to any specific model of interaction or entanglement with the universe and without perturbation and expansions and Markov approximations. The most widely known solution has been proposed by Ghirardi, Rimini and Weber (GRW) [3]. They modify phenomenologically the Schroedinger equation by adding nonlinear terms to the Hamiltonian so that the system, at Poisson distributed times, undergoes a sudden localization. Their model contains two unspecified parameters: the frequency and the spatial extent of localization. Other similar models have been proposed [4,5]. However, a common feature of the GRW model and of similar models is that *the energy of the system is not preserved*. Here, we look for an intrinsic mechanism for decoherence, which acts only on the coherence and not on the energy of the system.

A solution in this sense has been proposed by the author in 83 [6] assuming a finite difference Liouville equation with a time step $\tau$ (cronon) which has been related to the time-energy uncertainty relation. In ref. [7] it has been shown that this finite difference equation is equivalent to a semigroup master equation of the Lindblad form.

Recently, Milburn [8] has proposed a modification of the Liouville equation assuming that: "the system does not evolve continuously under unitary time evolution, but rather on a stochastic sequence of identical unitary transformations" according to a Poisson distribution. However, in the Milburn theory *time is a parameter*, as in standard quantum mechanics, and the Poisson distribution is assumed as the limit of a Bernoulli distribution.

In this paper we generalize the Liouville equation without any specific statistical assumption and without using any specific model for the interaction with the environment. We require only the semigroup property of the time evolution, dropping the unitarity requirement. We obtain an expression for the density operator which, a posteriori, can be interpreted *as if* the evolution time is not a fixed parameter but a *statistical variable* whose distribution is that of the waiting time of "line theory", i.e., a Γ-Poisson distribution. Our time distribution function contains two characteristic times which appear naturally in the theory as scaling times. They appears as the analog of time in the GRW localization of space. To be precise, time evolution appears as a series of random events in which $\tau_2$ is the average time step between two evolutions and $\tau_1$ is the time duration of each evolution. It is the same as saying that $\tau_2$ is the average interval between the arrival of two "clients" at the counter in a bank and $\tau_1$ is the time each "client" spends at the teller. Therefore, $\tau_1$ and $\tau_2$ have very different physical meaning and in general $\tau_1 \leq \tau_2$. Our time evolution law, for $\tau_1 = \tau_2 \to 0$ recovers the Liouville equation, whereas for $\tau_1$ and $\tau_2$ arbitrarily small but finite, results in a irreversible state reduction to the diagonal form in the energy representation. This state reduction has been postulated in ref.[9]. Here it is dynamically derived. Assuming $\tau = \tau_1 = \tau_2$ sufficiently small, we obtain the Milburn ME [8]. We suggest an application of our formalism to continuous measurement theory: $\tau_2^{-1}$ is the observation rate and $\tau_1$ is the time duration of each observation. Unlike the treatment of ref. [3], ours has energy as a constant of motion. Our evolution equation can be written in the form of a finite difference equation with time step $\tau_2$. Therefore, according to our formalism, one cannot give a continuos and instantaneous description of time evolution but only a description by "quantum jumps" with time steps given by the cronon $\tau_2$. From the finite difference equation we obtain a generalized Tam Mandelstam inequality which connects $\tau_1$ and $\tau_2$ with the time-energy uncertainty relation. Extending our formalism to a non Hamiltonian system we derive a generalization of the Scully-Lamb ME for laser and micromaser. Here $\tau_1$ and $\tau_2$ have a precise physical meaning because the micromaser model has been extensively investigated theoretically and experimentally to measure the quantum decoherence of an e.m. field prepared in a Schroedinger cat-like state, as one approaches the macroscopic limit (increasing the photon number) [10 − 13]. We apply our formalism to many examples which can be experimentally tested. In particular we derive an extra diffusion term in the position spread of a free particle; we describe localization of a free particle prepared in a Schroedinger cat-like state of two different positions. We demonstrate the existence of an intrinsic linewidth for a single mode e.m. field. Finally, we describe decoherence for a spin superposition state in a magnetic field and cancellation of EPR correlation under conditions in which they should not disappear according to standard quantum mechanics. Furthermore, intrinsic damping of Rabi oscillations in a two-level system, in agreement with recent experimental results [10, 14] is discussed.

## 2. The finite difference Liouville equation:

The unitary time evolution of a quantum system is generally described by the Liouville- von Neumann equation

(1) $$\frac{\partial \rho}{\partial t} = -iL\rho(t),$$

where $\rho$ is the density operator, L is the Liouvillian super-operator $L\rho \equiv \frac{1}{\hbar}[H,\rho]$, and H is the Hamiltonian. The formal solution of (1) can be written as

(2) $$\rho(t) = e^{-iLt}\rho_0 = e^{-\frac{i}{\hbar}Ht}\rho_0 e^{\frac{i}{\hbar}Ht}.$$

In (1) and (2) time t appears as a parameter, not an "observable" or a statistical variable as, for example, are position, momentum and H. Equation (1) and (2) can be expressed in the energy basis $H|n\rangle = E|n\rangle$ as:

(3) $$\dot{\rho}_{n,m} = -i\omega_{n,m}\rho_{n,m}$$

(4) $$\rho_{n,m}(t) = \rho_{n,m}(0)e^{-i\omega_{n,m}t}$$

where $\omega_{n,m} = (E_n - E_m)/\hbar$. The degenerate case can be included in this notation by assuming that the same states |n> belongs to the same eigenstate $E_n$. In the following treatment, in case of ambiguity with super-operators, one can refer to matrix elements replacing L with $\omega_{n,m}$ as it appears from Eqs.(2) and (4).

Let us define a generalized density operator $\bar{\rho}$ defined as

(5) $$\bar{\rho}(t) = \int_0^\infty dt' P(t,t')\rho(t'); \qquad t,t' \geq 0$$

where P(t,t') is a function still to be determined. In particular if $P(t,t') = \delta(t-t')$, $\overline{\rho}(t) = \rho(t)$, so that $\overline{\rho}(t)$ is a generalization of $\rho(t)$, if P is unspecified. Using Eq.(2) we can write

(6) $$\overline{\rho}(t) = \overline{V}(L,t)\rho(0)$$

where

(7) $$\overline{V}(L,t) = \int_0^\infty dt' P(t,t') e^{-iLt'}$$

We now determine P(t, t') imposing the following conditions:

i) $\quad \overline{\rho}(t) = \overline{\rho}^+(t) \geq 0; \qquad \text{Tr}\overline{\rho}(t) = 1$

ii) $\quad \overline{V}(t+t') = \overline{V}(t)\overline{V}(t')$

Condition (i) identifies $\overline{\rho}(t)$ as a density operator. Condition (ii) is the so called semigroup property which *ensures translational invariance* of the initial condition i.e., $\overline{\rho}(t+t') = \overline{V}(t+t')\rho(0) = \overline{V}(t)\overline{\rho}(t') = \overline{V}(t')\overline{\rho}(t)$. Note that (i) and (ii) are satisfied by the usual Liouville operator $V(t) = e^{-iLt}$. We just drop the request of Unitarity $\overline{V}^+ = \overline{V}^{-1}$. Condition (1), using (5) and taking the trace on both sides implies that

(8) $$\int_0^\infty dt' P(t,t') = 1; \qquad P(t,t') \geq 0$$

where we used $\text{Tr}\rho/t') = 1$. Therefore, Equation (8) defines P(t,t') as a probability distribution function; P(t,t')dt' can be read as the probability that the *random variable t* takes a value between t' and t'+dt'. Therefore, $\overline{\rho}(t)$ and $\overline{V}(t)$ appear as an average value respectively of $\rho(t)$ and $V(t)$. The semigroup property can be generally satisfied assuming

(9) $$\overline{V}(L,t) = [V_1(L)]^{t/\tau_2}$$

where $\tau_2$ is a scaling time. Note that also the Liouville operator $V(t) = e^{-iLt}$ can be written in the form (9) taking $V_1 = e^{-iL\tau_2}$ (unitary). We now determine the most general form of $V_1(L)$ compatible with the previous requirements. Let us use the $\Gamma$ function integral identity,

$$(10) \qquad (A+iB)^{-k} = \int_0^\infty d\lambda \frac{\lambda^{k-1}}{\Gamma(k)} e^{-A\lambda} e^{-iB\lambda} \ ;$$

which is valid for A>0 and k>0. Let us identify the following:

$$V_1^{-1} = A + iB \ ; \qquad k = t/\tau_2 \ ; \qquad \lambda = t'/\tau_1$$

where $\tau_1$ is a scaling time generally different from $\tau_2$. In this way by imposing the consistency of Eq.(10) with (7) for all L and the normalization condition (8), we obtain $B = L\tau_1$ and $A = 1$ so that

$$(11) \qquad \overline{\rho}(t) = \overline{V}(t)\rho(0) = \frac{1}{\left(1 + iL\tau_1\right)^{t/\tau_2}} \rho(0)$$

$$(12) \qquad \dot{\overline{\rho}}(t) = -\frac{1}{\tau_2} \ln(1 + iL\tau_1) \overline{\rho}(t)$$

and P(t, t') is the $\Gamma$ distribution function:

$$(13) \qquad P(t,t') = \frac{1}{\tau_1} \frac{e^{-t'/\tau_1}}{\Gamma(t/\tau_2)} \left(\frac{t'}{\tau_1}\right)^{(t/\tau_2)-1} \ ; \qquad (t,t' > 0)$$

This argument can be supported by considering P(t,t')=0 for t'<0 and using the unicity of the Fourier transform. Equations (11), (12) and (13) are the basic result of our paper. In particular, equation (11) provides a generalized form of the density operator and of the time evolution law which has been obtained just by requiring the density operator properties (i), the semigroup property (ii) and dropping the unitary requirement. Therefore, we are not adding new elements to the formalism of quantum mechanics but, on the contrary, we are reducing the basic assumptions. Taking $\tau_1 = \tau_2 \to 0$ in Eq.(11) one obtains the Liouville limit and $P(t,t') = \delta(t-t')$. However, as will be shown later, for $\tau_1$ and $\tau_2$ arbitrarily small but finite, as we shall see, one irreversibly approaches the diagonal form in the energy representation.

3. **The Finite Difference Equation:**

Equation (11) in the Liouville limit $\tau_1 = \tau_2 \to 0$ recovers again Eq.(1). When $\tau_1$ and $\tau_2$ are finite, the second order expansion of Eq.(11) gives

$$(14) \qquad \dot{\overline{\rho}} = -i\frac{L\tau_1}{\tau_2}\overline{\rho} - \frac{\tau_1^2}{2\tau_2}L^2\overline{\rho}$$

where $L^2\rho = [H,[H,\rho]]$. Equation (14) with $\tau_1 = \tau_2$ gives the well known "phase-destroying" Master Equation (ME), deduced by many authors using a reservoir interaction model [2] or specific statistical assumptions [8] and used in Quantum Non Demolition (QND) measurement theory [11]. It is straightforward to show from Eq.(11) that $\overline{\rho}(t)$ obeys the following finite difference equation

$$(15) \qquad \frac{\overline{\rho}(t) - \overline{\rho}(t-\tau_2)}{\tau_2} = -i\frac{\tau_1}{\tau_2}L\overline{\rho}(t) = -\frac{i}{\hbar}[\overline{H},\overline{\rho}(t)],$$

where $\overline{H} = H\tau_1/\tau_2$. Equation (15), for $\tau_1 = \tau_2 \to 0$, again gives the Liouville equation, whereas for $\tau_1 = \tau_2 = \tau$ it reduces to an equation proposed a long time ago [6] to describe irreversible state reduction to the diagonal form. The difference here is that now it has been derived under very general assumptions and two characteristic times appear, with very different physical meanings. In ref. [7], it has been shown that for $\tau_1 = \tau_2 = \tau$, Eq.(15) is equivalent to a semigroup ME of the Lindblad form. The same considerations apply here to Eq.(15), taking $\tau_2 = \tau$ and substituting H with $\overline{H}$. In this way Eq.(15) is formally identical to that considered in ref. [6].

Equation (15) shows a very important feature. If $\overline{\rho}(t)$ is a solution of (15), then $f(t)\overline{\rho}(t)$ is also a solution, provided $f(t+\tau_2) = f(t)$. Therefore, $\overline{\rho}(t)$ is uniquely determined only within the "cronon" $\tau_2$, i.e. for time intervals $t = k\tau_2$, with k integer. This fact implies a redefinition of $\rho(0)$ which in the standard description is the density operator determined at some instant t=0. This is clearly an artifact, because the possibility of an instantaneous measurement of a complete set of observables to determine $\rho(0)$ at the instant t=0 appears as a mathematical abstraction. Our finite interval description of time evolution, which follows from Eq.(15), appears much more realistic because $\rho(0)$ can be interpreted as the density operator determined in a cronon interval $\tau_2$. The evolution of Eq.(15) on the "time grid" gives $\rho$ at later time intervals, $k=t/\tau_2 \geq 1$, and can be parametrized as

(16) $$\overline{\rho}(k) - \overline{\rho}(k-1) = iL\tau_1 \overline{\rho}(k), \quad k \geq 1$$

or equivalently $\overline{\rho}(k+1) = (1+iL\tau_1)^{-1} \overline{\rho}(k) \approx e^{-iL\tau_1} \overline{\rho}(k)$, for $\tau_1$ small enough. It is *as if* time evolution occurs discontinuously in "quantum jumps" [12] spaced by intervals $\tau_2$ with each evolution occurring within a width $\tau_1$. Each "jump" is given by a unitary time evolution only if $\tau_1$ is small enough. Accordingly equations (7) and (11), for $t/\tau_2 = k \geq 1$, with $\Gamma(k) = (k-1)!$ can be written as:

(17) $$\overline{V}(k) = \int_0^\infty dt' P(k,t') e^{-iLt'} \; ;$$

and

(18) $$P(k,t') = \frac{1}{\tau_1} e^{-t'/\tau_1} \frac{(t'/\tau_1)^{k-1}}{(k-1)!},$$

where P(k, t') can be interpreted in two ways: i) as the well known Poisson distribution in k, or ii) as the $\Gamma$-distribution function in the continuous variable t'. Unlike previous treatments [8, 13] we adopt the latter. Equations (16) and (17) can be interpreted in terms of the *waiting time statistics* for k independent events. According to (17), *time evolution is made up of random "events"*, i.e., unitary time "evolution". The probability density for $k=t/\tau_2$ events to take place by a time t' is given by Eq.(18). In particular, $\tau_2$ is the average interval between two "events" ($\tau_2^{-1}$ is the rate) and $\tau_1$ is the time width of each event. We propose an interpretation of our formalism in terms of a continuos measurement theory: $\tau_2^{-1}$ is the *observation rate*, $\tau_1$ is the time width of each observation. As we shall see, this interpretation has a precise meaning in cavity QED measurements as well as in micromaser and laser theory [13]. In our formalism the interaction with the measurement apparatus is described by two characteristic time $\tau_1$ and $\tau_2$: this characterization is *intrinsic and model independent* in the sense that it does not depend on the way the measurement is carried out or on the detail on the measurement apparatus. Because the Hamiltonian is a constant of motion, our formalism can be applied, though not exclusively, to a QND measurement. According to this interpretation one obtains a dynamical definition of state reduction in agreement with the von Neumann state reduction. In fact, in general, in the energy representation Eq.(11) becomes

$$(19) \quad \overline{\rho}_{n,m}(t) = \frac{1}{\left(1+i\omega_{n,m}\tau_1\right)^{t/\tau_2}} \rho_{n,m}(0) = \frac{e^{-i\nu_{n,m}t}}{\left(\sqrt{1+\omega_{n,m}^2\tau_1^2}\right)^{t/\tau_2}} = e^{-i\gamma_{n,m}t}e^{-i\nu_{n,m}t}\rho_{n,m}(0)$$

where

$$(20) \quad \gamma_{n,m} = (1/2\tau_2)\ln(1+\omega_{n,m}^2\tau_1^2); \quad \nu_{n,m} = (1/\tau_2)\mathrm{arctg}\,\omega_{n,m}\tau_1.$$

Note that irreversibility is obtained also in the limit $\tau_1 \to 0$ and $\tau_2 \to 0$, provided that $\tau_1^2/\tau_2$ is finite, which describes an irreversible evolution. In general, the decoherence rates $\gamma_{n,m}$ increases as $\tau_2$ becomes smaller or $\tau_1$ becomes larger. Assuming, for simplicity, non degeneracy, for $E_n = E_m$, Therefore, $\omega_{n,m}=0$ and $\overline{\rho}_{n,n} = \rho_{n,n}(0)$, so that the energy is a constant of motion, whereas for $n \ne m$, $|\overline{\rho}_{n,m}| \to 0$ with a rate $\gamma_{n,m}$. Therefore, $\overline{\rho}(t) \to \sum_n \rho_{n,n}(0)|n\rangle\langle n|$ i.e., $\overline{\rho}$ approaches the stationary diagonal form. Then a pure state remains a pure state if and only if it is a stationary state; otherwise the system will evolve to a statistical mixture. Note that, at the microscopic limit, when the energy scale becomes very large, the decoherence time $1/\gamma$ becomes very short. The rate constants $\gamma_{n,m}$ and the frequencies $\nu_{n,m}$ have different behavior depending on whether $\omega_{n,m}\tau_1 \gg 1$ or whether $\omega_{n,m}\tau_1 \ll 1$. If $\omega_{n,m}\tau_1 \ll 1$ one has

$$(21) \quad \gamma \approx \frac{\omega^2 \tau_1^2}{2\tau_2} \quad \text{and} \quad \nu \approx \frac{\tau_1}{\tau_2}\omega$$

where we have omitted reference to m and n. This is also the result one obtains by the phase diffusing ME (14). Therefore, this ME is valid only if $\omega_{n,m}\tau_1 \ll 1$, for all n and m. This assumption can be made only if the spectrum is bounded so that it cannot be applied to a simple harmonic oscillator, as is usually assumed [11]. In the opposite limit, $\omega_{n,m}\tau_1 \gg 1$, one has

$$(22) \quad \gamma \approx \frac{\ln \omega \tau_1}{\tau_2} \quad \text{and} \quad \nu \approx \frac{\pi}{2\tau_2}.$$

Therefore, for large $\omega$ there is a slow logarithm dependence of the decoherence rate on the energy separation whereas $\nu$ becomes independent of $\omega$. We are now in a position to interpret the behavior of the frequencies $\omega_{n,m}$ as given by Eq.(21). If $\omega_{n,m}\tau_1 \ll 1$ the decoherence time $t_D=1/\gamma$ is much larger than $\tau_2$. Therefore, it takes many $\tau_2$ time step for decoherence to occur. Regarding the frequency, we note that using Eq.(21), $e^{i\nu t} = e^{i\omega \bar{t}}$, where $\bar{t} \equiv (t/\tau_2)\tau_1$ is the "effective evolution time" given by the number of evolution steps $t/\tau_2$, times the width of each evolution step, $\tau_1$. If $\omega_{n,m}\tau_1 \gg 1$ one has, from Eq.(22), a decoherence time $t_D = 1/\gamma = \tau_2/\ln(\omega\tau_1) \ll \tau_2$. This implies "immediate decoherence", i.e., von Newmann state reduction after the first evolution interval $\tau_2$. In this case of strong decoherence, the "frequency" loses its significance as it is smaller than the linewidth. We emphasize that the basic point of our treatment is that in our treatment *time appears as a statistical variable* with a distribution function P(t,t'), given by Eq.(13) and (18), which for k=t/$\tau_2$=1, is a simple exponential. However, for $t/\tau_2 \gg 1$, P(t,t') is a strongly peaked function with a mean value, <t'>, and dispersion, $\sigma$, given by

$$(23) \quad <t'> = \bar{t} \equiv (t/\tau_2)\tau_1,$$

$$(24) \quad \sigma = \sqrt{<t'^2> - <t'>^2} = \tau_1\sqrt{t/\tau_2}$$

Note that <t'> does not coincide with t but with the effective "evolution time" $\bar{t}$. The dispersion σ scales as $\sqrt{t}$, like in a diffusion process. According to the previous interpretation, the dispersion of σ appears as the dispersion due to k=t/$\tau_2$ statistically independent "events" times the width of each event $\tau_1$. For k = 1, σ assumes its minimum value $\tau_1$. Therefore, $\tau_1$ appears as an "inner time" [15], i.e., the intrinsic minimum uncertainty of the time evolution. The relative dispersion, $\sigma/<t'> = (t/\tau_2)^{-1/2}$, goes to zero as the number of time steps, t/$\tau_2$, goes to infinity. Furthermore, for t/$\tau_2$ >>1, P(t, t') can be approximated by a Gaussian in t' with mean value given by Eq.(23) and dispersion given by Eq.(24).

## 4. Comparison with previous work

The difference between our theory and previous descriptions of intrinsic decoherence [8] can be summarized as follows: In ref. [8] it is assumed that i) the system evolves under a random sequences of identical unitary transformations, and ii) the probability of n transformations in a time t is given by a Poisson distribution so that (using our notation)

$$(25) \qquad \bar{\rho}(t) = \sum_{n=0}^{\infty} p(n,t) e^{-inL\tau_1} \rho(0)$$

where $p(n,t) = \frac{(t/\tau_2)^n}{n!} e^{-t/\tau_2}$. In ref.[8] $\tau_1 = \tau_2 = \gamma^{-1}$ is interpreted as a "fundamental time of the universe". Summing up the series (25) one obtains $\bar{\rho}(t) = \exp\left(\frac{t}{\tau_2}\left(e^{-iL\tau_1} - 1\right)\right) \rho(0)$ so that

$$(26) \qquad \dot{\bar{\rho}}(t) = \frac{1}{\tau_2}\left(e^{-iL\tau_1} - 1\right)\bar{\rho}(t) = \frac{1}{\tau_2}\left(e^{-iH\tau_1/\hbar}\bar{\rho}(t)e^{iH\tau_1/\hbar} - \bar{\rho}(t)\right)$$

Equation (26), with $\tau_1 = \tau_2 = \gamma^{-1}$ is the basic result of ref. [8]. If one applies Eq.(26) with this assumption to calculate the average value of the annihilation operator of an harmonic oscillator, one obtains easily $<\dot{a}> = \gamma\left(e^{-i\omega/\gamma} - 1\right)<a>$. From this equation Milburn [8] infers that there are particular frequencies ω=2nπγ where the oscillator is "frozen", i.e., $<\dot{a}> = 0$. As we shall see this unusual behavior does not occur in our treatment because we always have damped amplitude for all frequencies. We note that the same method has been adopted in Laser-Micromaser and cavity QED theories [13], where the formal substitution:

$$(27) \quad e^{-iL\tau_1} \Rightarrow M(\tau_1) \quad ; \quad L \Rightarrow (i/\tau_1)\ln M$$

have been made in Eq.(25). Here M is a non unitary operator well known in laser-micromaser theory described in ref.[13]. In this way Eq.(26) becomes the well known Scully-Lamb ME

$$(28) \quad \dot{\bar{\rho}} = (1/\tau_2)(M(\tau_1)-1)\bar{\rho}$$

where $\tau_2^{-1} = r$ is the rate of injection of the atoms and $\tau_1$ is the atomic interaction time. The basic differences between our approach and Milburn approach are:

i) we do not make any statistical assumption: randomness in time evolution appears naturally as an interpretation of our results, i.e., of Eq.(7) and (13). In particular the Poisson-$\Gamma$ distribution, Eq.(13), is not assumed but derived.

ii) the choice $\tau_1 = \tau_2$ implies a severe restriction on the statistical interpretation of the theory. In fact, in our treatment, as well as in the Scully-Lamb theory, $\tau_1$ and $\tau_2$ have very different physical meanings.

iii) Equation (25) assumes that the number of evolution transformations in a time t is a random variable and t is just a parameter as in the Schroedinger equation or in the Liouville equation. In our theory *the waiting time* t for n evolutions to occur is the random variable.

The two view points appear similar but are basically different. In fact, they lead to different results, as one can see by comparing Eq.(26) and Eq.(12). The first can be derived from the second as follows. Using the integral identity:

$$(29) \quad \ln(1+ix) = \int_0^\infty d\lambda \frac{e^{-\lambda}}{\lambda}\left(1-e^{-i\lambda x}\right)$$

and taking $x = L\tau_1$ and $\lambda = t'/\tau_1$, Eq.(12) can be written in the integral form

(30) $$\dot{\bar{\rho}}(t) = \frac{\tau_1}{\tau_2}\int_0^\infty dt'\left(\frac{1}{\tau_1}e^{-t'/\tau_1}\right)\left[\frac{e^{-iLt'}-1}{t'}\right]\bar{\rho}(t)$$

This equation is completely equivalent to Eq.(12). It can be shown that if the term in square brackets is slowly varying on a time scale $\tau_1$, it can be taken out of the integral with t'= $\tau_1$. Because the term in the round brackets is normalized, one obtains the Milburn ME (26). Therefore, Eq.(26) can be obtained as an approximation of Eq.(12), if $\tau_1 = \tau_2 = \gamma^{-1}$ is small enough. In the case of an harmonic oscillator this implies the assumption $\omega/\gamma \ll 1$, which is inconsistent with the condition for freezing $\omega/\gamma = 2n\pi$. Note that by making the substitution (27) in Eq.(12) and (30) one obtains a generalized laser-micromaser ME:

(31) $$\dot{\bar{\rho}}(t) = \frac{\tau_1}{\tau_2}\int_0^\infty dt'\frac{1}{\tau_1}e^{-t'/\tau_1}\left[\frac{M^{t'/\tau_1}-1}{t'}\right]\bar{\rho}(t) = -\frac{1}{\tau_2}\ln(1-\ln M)\bar{\rho}(t).$$

Again if $\tau_1$ is small enough, Eq.(31) reduces to the Scully-Lamb ME (28). Therefore, Eq.(31), with no restrictions on $\tau_1$, appears as a generalization of Eq.(28). As already mentioned, in the context of cavity QED [13] the two characteristic times have a very precise physical meaning. It is interesting to note that the finite difference equation (15), with the substitution (27), reads $\frac{\bar{\rho}(t)-\bar{\rho}(t-\tau_2)}{\tau_2} = \left(\frac{1}{\tau_2}\ln M\right)\bar{\rho}(t)$. Replacing the RHS with the derivative, one obtains

(32) $$\frac{d\bar{\rho}}{dt} = \left(\frac{1}{\tau_2}\ln M\right)\bar{\rho},$$

which is the equation of a micromaser with regular injection [13]. Finally, dissipation can be taken into account by adding the proper dissipative term of the Lindblad form to Eq.(12) or (31). The possibility of applying our formalism to cavity QED is of particular relevance to the quantum theory of measurement. Cavity QED has been extensively investigated both theoretically and experimentally [10,13].

## 5. Generalized Tam Mandelstam Relation:

From the finite difference equation (15), and defining the mean value of an observable A in the usual way, $\overline{A} = \text{Tr}\overline{\rho}A$, we obtain

$$\text{(33)} \qquad \frac{\overline{A}(t) - \overline{A}(t - \tau_2)}{\tau_1} = -\frac{i}{\hbar}\overline{[A, H]}.$$

Let us apply Eq.(33) to the one dimensional motion of a particle with Hamiltonian $H = p^2/(2m) + V(x)$. Taking A=x and A=$p_x$ we have $(\overline{x}(t) - \overline{x}(t - \tau_2))/\tau_1 = \overline{p}_x(t)$ and $(\overline{p}_x(t) - \overline{p}_x(t - \tau_2))/\tau_1 = \overline{F(x)}$, where $F(x) = -dV/dx$. This is the finite difference version of the Herenfest theorem. The usual form is regained in the continuos limit $\tau_1 = \tau_2 \to 0$, whereas the classical limit is $\overline{F(x)} \approx F(\overline{x})$. Note that the two limits are independent. Therefore, taking only the classical limit one obtains *finite difference Hamiltonian equations* for $\overline{x}$ and $\overline{p}_x$.

We now derive a generalized Tam Mandelstam relation. Using the general uncertainty relation for A and H we can write

$$\text{(34)} \qquad \sigma(A)\sigma(H) \geq \frac{1}{2}\left|\overline{[A, H]}\right| = \frac{\hbar}{2}\frac{|\Delta\overline{A}|}{\tau_1}$$

where $\Delta\overline{A} = \overline{A}(t) - \overline{A}(t - \tau_2)$ and we have used Eq.(33). Equation (34) can be written in the form

$$\text{(35)} \qquad \tau_A \sigma(H) \geq \hbar/2 \ ; \qquad \tau_A \equiv \frac{\sigma(A)}{|\Delta\overline{A}|/\tau_1}.$$

This equation appears as a generalized TM inequality and reduces to the usual form in the limit $\tau_1 = \tau_2 \to 0$. In fact in this way one obtains $\tau_A = \sigma(A)/|\dot{\overline{A}}|$. However, going back to Eq. (34) one obtains the inequality

$$\text{(36)} \qquad \frac{|\Delta\overline{A}|}{\sigma(A)} \leq \frac{\tau_1}{\tau_E}$$

where $\tau_E = \hbar/2\sigma(H)$ is the intrinsic inner time [15] of the system. Because inequality (36) is valid for any observable, we can conclude that if $\tau_1 \leq \tau_E$ then $|\Delta \overline{A}| \leq \sigma(A)$ necessarily follows. Therefore, no appreciable variation occurs for any observable in the cronon time $\tau_2$. We can say that if $\tau_1 \leq \tau_E$ one has a quasi continuous evolution of the system even using the discrete time description. On the other hand, Eq. (36) states that if $\tau_2$ is such that $|\Delta \overline{A}| \geq \sigma(A)$, it follows that one must have $\tau_1 \geq \tau_E$. Therefore, the ratio $\tau_1/\tau_E$ rules the rate of change of the state of the system within the time interval $\tau_2$ between two evolutions. The choice $\tau_1 = \tau_E = \hbar/2\sigma(H)$ corresponds to the maximum possible value of $\tau_1$ which guarantees the quasi continuous evolution which is commonly observed. This choice is clearly meaningless in case of a non Hamiltonian system, as in Eq.(31). Furthermore, from $\tau_2 \geq \tau_1 = \tau_E$, one would obtain $\tau_2 \sigma(H) \geq \hbar/2$, giving a precise and intrinsic meaning to the time-energy uncertainty relation.

## 6. Calculation of Physical Quantities:

Equation (5) allows one to calculate all physical quantities by just performing a time integral of the usual expression. Let us give some examples: The mean value $\overline{A}(t) = \text{Tr}\overline{\rho}(t)A$ can be easily obtained using Eq.(5) as

(37) $$\overline{A}(t) = \int_0^\infty dt' P(t,t') <A>_{t'}$$

where

(38) $$<A>_t = \text{Tr}\rho(t)A$$

is the usual expectation value. Therefore, a constant of motion remains a constant of motion, whereas oscillating quantities, as $e^{i\omega t}$, are damped with a rate constant $\gamma = (1/2\tau_2)\ln(1+\omega^2\tau_1^2)$ similarly to $\rho_{n,m}$ as given by Eq.(19). Furthermore, from Eq.(5),

(39) $$\langle x|\overline{\rho}(t)|x'\rangle = \int_0^\infty dt' P(t,t')\langle x|\rho(t')|x'\rangle.$$

The same relation holds for matrix elements in any representation. However, Eq.(39) is of particular relevance. In fact, for x = x' it gives the position probability density, $\bar{P}(x,t)$. In particular, if the initial state is a pure state, $|\psi_0\rangle\langle\psi_0|$, then $\langle x|\rho(t')|x\rangle = |\psi(x,t')|^2$. Furthermore, because the Wigner function is related to the Fourier transform of $\langle x|\rho(t)|x'\rangle$ it can be obtained using the same integral relation as in Eq.(39). It can be easily shown that the same relation is valid for the Glauber P-function. We now make some testable application of our formalism.

## 7. Harmonic oscillator:

In this case the photon number is a constant of motion. On the contrary the amplitude <a> goes to zero. In fact, because $<a>_t = <a>_0 e^{-i\omega t}$, from (37) and (8), one obtains $\bar{a}(t) = \frac{1}{(1+i\omega\tau_1)^{t/\tau_2}} <a>_0 = <a>_0 e^{-\gamma t}e^{-i\nu t}$, where $\gamma = \frac{1}{2\tau_2}\ln(1+\omega^2\tau_1^2)$ and $\nu = \frac{1}{\tau_2}\text{arctg}\,\omega\tau_1$. Therefore, $|\bar{a}(t)|$ goes to zero as t goes to infinity. The behavior of $\gamma$ and $\nu$ is formally the same as that of $\gamma_{n,m}$ and $\nu_{n,m}$ described in Eq.(20). Therefore, the same discussions apply just abolishing the index n,m. Moreover, in general, if one assumes that the initial state is a coherent state, $|\alpha_0\rangle$, the state at time t will be the coherent state $|\alpha(t)\rangle$ with $\alpha(t) = \alpha_0 e^{-i\omega t}$. Therefore,

$$(40) \quad \rho(t) = \sum_{n,m} e^{-\bar{n}} \frac{\alpha_0^n (\alpha_0^*)^m}{\sqrt{n!m!}} e^{i\omega(n-m)t} |n\rangle\langle m|$$

where $\bar{n} = |\alpha_0|^2$. The oscillatory behavior of the diagonal elements implies that as $t \to \infty$,

$\bar{\rho}(t) \to \sum_n e^{-\bar{n}}(\bar{n})^n / n!$. This happens because $|\bar{\rho}_{n,m}(t)| = e^{-\gamma_{n,m}t}$ with $\gamma_{n,m} = (1/\tau_2)\ln[1+\omega^2(n-m)^2\tau_1^2]$, where Eq.(21) has been used. Note that the decoherence rates increase as one increases the distance between energy levels. For n and m such that $\omega|n-m|\tau_1 << 1$, expanding the logarithm to the first order, $\gamma_{n,m} \propto |n-m|^2$, as has been already derived by other authors [11] with the ME (20) whereas if

|n-m| is large enough $\gamma_{n,m}$ has only a logarithmic dependence on |n-m| and the ME (14) does not apply.

The diagonal density operator (40) has been introduced by Glauber [16] to describe the steady-state of a laser well above threshold using a random phase assumption for $\alpha_0$. Here, the approach to the diagonal form is dynamically obtained even if the phase of $\alpha_0$ is perfectly determined. Note that the amplitude damping should be experimentally observable by proper homodyne detection or, more simply, by measuring the linewidth. This intrinsic linewidth $\gamma$ should be observable only if it dominates on all other linewidths, i.e., if $\tau_1$ is large enough and $\tau_2$ is small enough. In the opposite case one obtains a lower limit for $\tau_2$ and an upper limit for $\tau_1$.

## 8. Free particle spread:

Let us assume to have a free particle in a minimum uncertainty packet, $\langle x|\rho|x\rangle = |\psi(x,t)|^2 =$ $\left(1/\sqrt{2\pi}\sigma(t)\right)e^{-(x-<x>_t)^2/2\sigma^2(t)}$, where $<x>_t = x_0 + vt$, $\sigma^2(t) = \sigma_x^2 + \sigma_v^2 t^2$ with $v = p/m$ and $\sigma_v = \sigma_p/m$. Using the integral (39) with x=x' and a Gaussian approximation for P(t,t') one obtains again a Gaussian for $\overline{P}(x,t)$ centered on $\overline{x}(t) = x_0 + v\overline{t}$ and spread

$$(41) \qquad \overline{\sigma}_t^2 \approx \sigma_x^2 + \sigma_v^2 \overline{t}^2 + v^2 \tau_1 \overline{t}$$

where $\overline{t}$ is the reduced time $\overline{t} = t\tau_1/\tau_2$. If $\tau_1 = \tau_2$, this expression coincides with the one derived in ref. [8]. Assuming $v >> \sigma_v$, this term is dominant, if $t < (v/\sigma_v)^2 \tau_2$. So, if $v/\sigma_v$ and $\tau_2$ are large enough, the diffusion term should be observable.

## 9. Space Localization in a Superposition State:

Let us consider a free particle prepared in a superposition state of two minimum uncertainty wave packets, centered at different positions $x_1$, and $x_2$. The wave function at time t is given by

$$(42) \qquad \psi(x,t) = \left(1/\sqrt{2}\right)\left(\psi_1(x,t) + \psi_2(x,t)\right)$$

where

$$(43) \quad \psi_j(x,t) = \frac{1}{\left(\sqrt{2\pi}\sigma_x\right)^{1/2}} e^{-(x-x_j)^2/4\sigma_x^2} \left(1 - i\frac{\sigma_v}{\sigma_x} t\right) \quad j=1,2.$$

Here, for simplicity, we have neglected the Schroedinger spread, by assuming $\sigma_v$ to be small enough. This is certainly the case at the macroscopic limit, because $\sigma_v = \hbar/2m\sigma_x$, where m is the mass of the particle. From Eq.(42) we have:

$$(44) \quad P(x,t) = |\psi(x,t)|^2 = (1/2)|\psi_1(x)|^2 + (1/2)|\psi_2(x)|^2 + \text{Int}$$

where $\text{Int} = |\psi_1(x)||\psi_2(x)|\cos\omega t$, with

$$(45) \quad \omega = \sigma_v x D / 2\sigma_x^3 \;;$$

with the assumption that $x_1 = -x_2 = D/2$. Therefore, in standard quantum mechanics the interference term is oscillating in time. In contrast, in our formalism the interference disappears because, using Eq.(37) and (8), the average of $\cos(\omega t)$ is exponentially damped at a rate given by

$$(46) \quad \gamma = (1/2\tau_2)\ln(1+\omega^2\tau_1^2),$$

where $\omega$ is given by Eq.(45). The interference disappears in a decoherence time $t_D = \gamma^{-1}$. Note that if $\omega\tau_1 \ll 1$ one can expand the logarithm to the first order, obtaining a decoherence rate proportional to $\omega^2$, i.e., to the square of the distance between the centers of the two packets, in agreement with previous treatments. The interaction with the environment and/or with the measuring apparatus appears quite naturally in our formalism via the two characteristic times without specifying any interaction model. If we use the uncertainty principle to eliminate $\sigma_x$, Eq.(45) can be written as

$$(47) \quad \omega = 16mE^2 Dx / \hbar^3,$$

where $E = (1/2)m<v^2>$. This expression, together with Eq.(45), clearly shows that at the macroscopic limit, when m and E are very large, or at the classical limit $\hbar \to 0$, the decoherence rate becomes extremely fast. Let us remember the fundamental and mysterious meaning of the interference term. Its presence implies that to find the particle around $x_1$ *or* around $x_2$ are *not* *mutually exclusive events.* Otherwise, this probability would be, by definition, the sum of the probabilities. Therefore, the interference term, independently of its magnitude, says, roughly speaking that the particle is in a "schizophrenic" superposition of here and there. This is substantially the Schroedinger cat paradox, if $x_1$ and $x_2$ are replaced by "dead" or "alive". Quoting Feynman, "interference is *the only* mysteries of quantum mechanics" [17]. We call decoherence or state reduction, the suppression of the interference term so that the particle can be found at $x_1$ *or* $x_2$ with 50% classical probability, but without being able to predict where. We think that this is the "best" localization quantum mechanics can describe without strongly changing its formalism. Other authors [3] introduce non linear terms in the Schroedinger equation to have not only the suppression of the interference term but also the reduction to a single wave packet which, in the Copenhagen interpretation should take place only when one really observes the particle. Accordingly, only the suppression of the interference term should be intrinsic. Many authors [1,18] describe decoherence as due to dissipation, which is like saying that the particle is here or there because it is going to be stopped by friction or the cat cannot be dead and alive because it is going to dissipate its energy in the universe and die. It is clear that dissipation is *sufficient* for decoherence. But it is not clear that it is *necessary* for decoherence. The interference does not disappear at x=0 because there the interference term is stationary. For this reason our decoherence mechanism is not present in the case of the stationary two slits interference experiment between particles having the same energy or momentum. In our case the time dependence is due to the finite velocity spread. The decay rate, even if it is zero exactly at x=0, is different from zero in a region whose extent goes to zero at the macroscopic limit. Since the probability is the integral of the probability density, the interference term goes to zero in a finite region around x=0.

10. **Decoherence in two level systems:**

Let us assume a spin system of neutral particles injected one at a time with velocity v, in a constant magnetic field $B_0$ of extent L in the z direction. The particles are prepared in a eigenstate of $\sigma_x$, $|\psi_0\rangle = \frac{1}{\sqrt{2}}(|+\rangle + |-\rangle)$. Therefore, $|\psi\rangle_t = \frac{1}{\sqrt{2}}(e^{i\omega t/2}|+\rangle + e^{-i\omega t/2}|-\rangle)$. Here $\omega_0 = \mu B_0/\hbar$ is the Larmor frequency and $\mu$ is the Bohr magneton. The density operator at time t reads $\rho(t) = |\Psi\rangle\langle\Psi| = \frac{1}{2}(\rho_D + e^{i\omega_0 t}|+\rangle\langle-| + h.c)$, where $\rho_D = (1/2)[|+\rangle\langle+| + |-\rangle\langle-|]$ is the diagonal part. Note that for $\omega_0 t = 2n\pi$ one has again the initial pure state. However, using our formalism the off-diagonal elements will be damped as $e^{-\gamma t}$ where

(48) $$\gamma = (1/2\tau_2)\ln(1+\omega_0^2\tau_1^2)$$

and t = L/v is the transit time in the magnetic field. Let us now insert, after the region L a Stern-Gerlach apparatus in the x direction and let us take $\omega_0 L/v = 2n\pi$ with n arbitrarily large. According to standard quantum mechanics all particles would go only in one direction. However, if $\gamma L/v$ is large enough, our decoherence will take place, i.e., 50% of the particle will go both directions. The non observation of this decoherence would put a upper limit for $\tau_1$ and a lower limit for $\tau_2$.

11. **Intrinsic Damping of Rabi Oscillations:**

The same considerations can be applied to a system of two level atoms injected one at the time in a high Q resonant cavity and prepared in a Rydberg state so that the atomic and cavity decay time and the intrinsic decoherence are very long. This is the experimental situation in ref. [10]. If the field is in a n-Fock state and the atoms are injected in the excited state, the atoms will oscillate between the upper and lower state so that the population difference d will oscillate as $d = \cos\Omega t$, where $\Omega = g\sqrt{n+1}$ is the Rabi frequency and g is the one photon Rabi frequency accordingly with the Jaynes-Cummings atom-field interaction Hamiltonian model. In our formalism these oscillations will be damped with a rate constant, $\gamma = (1/2\tau_2)\ln(1+\Omega^2\tau_1^2)$. Therefore, if $\gamma t = \gamma L/v >> 1$, the population difference will approach the steady state value d=0 so that the population of the upper level will approach the value 0.5. This behavior has been indeed observed in ref.[10] also for the vacuum state where dissipation losses are ineffective and in ref. [14] for a n-Fock state, in a different experimental situation. In particular, in ref. [14], it has been observed an increase of the damping rate γ with n, in "qualitative" agreement with $(n+1)^{0.7}$. Let us point out that our damping rate has a logarithmic dependence on (n+1), which, if $\tau_1$ is small enough, goes like (n+1). Whether this damping is due to experimental problems, as suggested by the authors, or can be attributed to our intrinsic decoherence mechanism deserves further experimental and theoretical investigation. Obviously the same intrinsic damping can be predicted for the Rabi oscillations in a NMR configuration.

**12. Cancellation of EPR correlation:**

Let us assume two spin-1/2 neutral particles, say 1 and 2, traveling in opposite directions with velocity v and prepared in the singlet entangled state $|\psi_0\rangle = \frac{1}{\sqrt{2}}(|+_1,-_2\rangle - |-_1,+_2\rangle)$ with a constant magnetic field $B_0$ in the z direction acting on the path of particle 1 for a length L. We have $|\psi\rangle_t = \frac{1}{\sqrt{2}}(e^{i\omega_0 t/2}|+,-\rangle - e^{-i\omega_0 t/2}|-,+\rangle)$, where $\omega_0$ is the Larmor frequency. The associated density operator reads

(49) $$\rho(t) = |\psi(t)\rangle\langle\psi(t)| = \left(\rho_D - \frac{1}{2}e^{i\omega_0 t}|+,-\rangle\langle-,+| - h.c\right)$$

where $\rho_D$ is the diagonal part. Note that for $\omega_0 L/v = 2n\pi$ one has again the initial singlet state. Therefore, the presence of a e.m. field would become ineffective regarding EPR correlation under these conditions. However, in our formalism the oscillating of the off-diagonal terms disappears exponentially with γ given by Eq.(48). Therefore, if γL/v >>1 the EPR correlation should disappear even if $\omega_0 L/v = 2n\pi$. Any evidence of correlations smaller than expected in the conditions described above, would be a strong support of our theory.

**Conclusion:**

Using Anton Zeilinger [19] words: "When investigating various interpretations of quantum mechanics one notices that each interpretation contains an element which escapes a complete and full description. This element is always associated with the stochasticity of the individual event … It is suggested that the objective randomness of the individual quantum event is a necessity of a description of the word…. Yet it is also highly recommended to follow the guidance of the Copenhagen interpretation, that is, not to make any unnecessary assumption."

In this spirit we have generalized the Liouville equation without introducing new assumptions to quantum mechanics. On the contrary, we have reduced the basic axioms of quantum mechanics dropping the unitarity condition and maintaining only the semigroup property of the time evolution operator. The new equation describes intrinsic decoherence giving irreversible state reduction to the diagonal form in the energy basis so it can be used to describe quantum non demolition measurements. In our treatment time evolution appears, a posteriori as a statistical variable obeying "the waiting time statistics" whereas in standard quantum mechanics as well as in previous descriptions of intrinsic decoherence time is just a parameter. Our evolution equation can be written as a finite difference equation, which looks *as if* time evolution is a random process with two characteristic times $\tau_1$ and $\tau_2$. The second time is the average interval between two evolution and $\tau_1$ is the time width of each evolution. For $\tau_1 = \tau_2 = \tau$ one recovers a previously proposed finite difference equation [6], which for $\tau$ small enough and in a particular limit, gives the ME for intrinsic decoherence proposed in ref. [8]. A generalized Tam Mandelstam inequality has been derived, which suggests a relation between $\tau_1$ and the "inner time" of the system, $\hbar/2\sigma(H)$ [15]. This relation guarantees a quasi continuous evolution even with a finite difference equation. We have shown that our equation implies localization of a free particle prepared in a Schroedinger cat-like superposition of states with different position as the macroscopic limit is approached. The relation of our theory with micromaser theory and experiments have also been discussed. Some testable examples have been described. In particular, we predict the damping of Rabi oscillations, when all dissipation mechanisms appears to be ineffective, in agreement with the experimental result of Ref. [10] and [14].

**Acknowledgments**

We must acknowledge that our work started in 1983 stimulated by Caldirola's pioneering work on a finite difference Schroedinger equation to describe the quantum theory of the electron [20]. We are indebted to L. Davidovich and N. Zagury for their hospitality at the Federal University of Rio de Janeiro where this work has been partially developed under their continuous stimulation, suggestions and criticisms. We are grateful to S. Osnaghi for his continuous collaboration, L. De Salvo for her assistance and to D. A. Jaroszynski for reading, improving the manuscript and fruitful discussions.

**References**

1. Giulini D., Joos E., Kiefer C., Kupsch J., Stamatescu I.-O., Zeh H. D., Decoherence and the Appearence of a Classical World in Quantum Theory (Springer, 1996)

2. Omnès R., Phys. Rev. A*56*, 3383 (1997) and references therein.

3. Ghirardi G. C., Rimini A., Weber T., Phys. Rev. D*34*, 470 (1986); Phys. Rev. D*36*, 3287 (1987) and references therein.

4. Diosi L., Phys. Rev. A*40*, 1165 (1989); Ghirardi G. C., Pearle P., and Rimini A., ibidem, *42*, 78 (1999).

5. Percival I. C., Proc. Roy. Soc., 451, 503 (1995); Bell J. S., Physics Word, 3, 33 (1990).

6. Bonifacio R., Lettere al Nuovo Cimento, *37*, 481 (1983).

7. Ghirardi G. C., Weber T., Lett. Nuovo Cimento, *39*, 157 (1984).

8. Milburn G. J., Phys. Rev. A*44*, R5401 (1991); Moya-Cessa H., Buzek V., Kim M. S., and Knight P. L., Phys. Rev. A*48*, 3900 (1993).

9. Bedford D., Wang D., Il Nuovo Cimento B*37*, 55 (1977).

10. Brune M., Hagley E., Dreyer J., Maitre X., Maali A., Wunderlich C. Raimond J., Haroche S., Phys. Rev. Lett. *77*, 4887 (1996)..

11. Milburn G. J., Walls D. F., Phys. Rev. A*38*, 1087-1090 (1988).

12. Caves C. M., Milburn G. J., Phys. Rev. D*36*, 3543 (1987); Percival I. C., Proc. Royal Soc. 451, 503 (1995) and references therein.

13. Davidovich L., Rev. Mod. Phys. *68*, 127 (1988) and references therein.

14. Meekhof D. M., Monroe C., King B. E., Itano W. M., Wineland D. J., Phys. Rev. Lett., *76*, 1796 (1996). A different interpretation of this experiment has been given by Schneider S., Milburn G. J., Phys. Rev. *A 57*, 3748 (1998); Murao M., Knight P. L., Phys. Rev. *A 58*, 663 (1998).

15. Aharonov Y., Bohm D., Phys. Rev. *122*, 1649 (1961).

16. Glauber R., Les Houches 1964. Quantum Optics and Electronics (eds. C. De Witt, A. Bladin, C. Cohen-Tannoudji) (Gordon and Breach, 1965).

17. Feynman R., Leighton R., Sands M., The Feynman Lectures on Physics vol. III (Addison Wisley, Reading, 1965)

18. Zureg W. H., Physics Today, p.36, October (1991) and references therein.

19. Zeilinger A., "Vastakohtien todellisuus", Festschrift for Laurikainen K. V. (eds. U. Ketvel et al.), Helsinki University Press, 1996.

20. Caldirola P., Nuovo Cimento A, *45*, 549 (1978); Caldirola P., Riv. Nuovo Cimento, *2*, n. 13 (1979); Caldirola P. and Bonifacio R., Lett. Nuovo Cimento, *33*, 197 (1982).